\documentclass[aps,twocolumn,prl,showpacs,color,psfig,epsf]{revtex4}
\usepackage{amsmath}
\usepackage{color}
\usepackage{amsfonts}
\usepackage{epsf}
\usepackage{graphicx}
\begin{document}

\title{Melting a stretched DNA}

\author{D. Marenduzzo$^1$, A. Maritan$^{2,3}$,
E. Orlandini$^{2,3}$, F. Seno$^{2,3}$, A. Trovato$^2$}

\affiliation{
$^1$ SUPA, School of Physics, University of Edinburgh, Mayfield Road,
Edinburgh EH9 3JZ, Scotland \\
$^2$ Dipartimento di Fisica, Universita' di Padova,
and CNISM, Unit\`a di Padova, Via Marzolo 8, 35131 Padova, Italy \\
$^3$ INFN, Sezione di Padova, Via Marzolo 8, 35131 Padova, Italy }

\begin{abstract}
We study the melting of a double stranded DNA in the presence of stretching 
forces, via 3D Monte-Carlo simulations, exactly solvable models and heuristic
arguments. The resulting force-temperature phase diagram is 
dramatically different for the cases where the force is applied 
to only one strand or to both. Different assumptions on the monomer
size of single and double stranded DNA lead to opposite conclusions as to 
whether DNA melts or not as it overstretches.
\end{abstract}

\maketitle

Single molecule experiments have by now enriched our knowledge of
biopolymer physics at the nanoscale\cite{Marko,Essevaz,rouzina}. DNA
may be grabbed, twisted and pulled with atomic force microscopes, laser 
tweezers etc., and in this way a number of its physical, chemical and elastic
properties may be measured.
Arguably the best characterised single molecule experiment is the adiabatic 
stretching of single stranded DNA (ssDNA) and of double stranded DNA (dsDNA),
the double helical B-form. 
In most cases, the resulting force-elongation curves may be 
described by simple statistical mechanics models, e.g. the worm-like 
or the freely-jointed chain. These inextensible polymer models 
work at relatively low force, and for chain length larger than the persistence 
length, $\sim 50$ nm ($150$ base-pairs (bp)) for dsDNA and $\sim 2$ nm ($4$ bp) for ssDNA. 
Still, this classic experiments has prompted several other questions, some
still outstanding. Most relevant for us is the fact that, at 
about $\sim 65$ pN, dsDNA elongates to about 1.7 times the B-DNA
contour length: this regime is usually referred to as {\em overstretching}. Is 
overstretched DNA melted, ssDNA, or is it a different form of dsDNA,  
named S-DNA~\cite{rouzina,Clausen-Schaumann00}? These two competing pictures 
are not easily discriminated, since the contour lengths of S-DNA and
ssDNA at the overstretching transition are similar ($0.56$-$0.58$ nm per
bp). A further complication is given by the possibility of pulling one
or both strands -- experiments with laser tweezers may often not be able
to tell what is happening~\cite{rouzina}, as one strand may fall off the
other due to the presence of nicks breaking the DNA backbone,
{\it unpeeling}~\cite{Cocco}.
Are the two ensembles equivalent, or any similar?

Theories of DNA overstretching
are often phenomenological, determining parameters
from fitting to data~\cite{Cocco,Nelson,Geissler}.
On the other hand, a lot of theoretical work focused on
simpler models to find the temperature-force ($T-f$) phase diagram of
biopolymers perturbed by external forces  -- see
e.g.~\cite{dna_unzipping1,dna_unzipping3,stretching,kumar,somen}.
DNA overstretching was only recently tackled within similarly
simplified models, by assuming that it amounts to melting ds
to ssDNA~\cite{Hanke,rudnick}. On the basis of a continuum
model~\cite{Hanke}, it was proposed that when one strand is being
pulled, there should be a ``reentrance'' at low $T$, similar to that 
found for DNA unzipping~\cite{dna_unzipping3}. Another interesting
debated question is whether the ds phase is destabilised or not by a
stretching force~\cite{rouzina,Hanke}. A final prediction is that the
melting transition, which is first order at $f=0$,
should become second order at $f\ne 0$\cite{Hanke,rudnick}.
 
Here we will be concerned with the melting transition of a model of
dsDNA anchored on one side and subject to external stretching forces,
$\mathbf{f}_1$ and $\mathbf{f}_2$ on the two strands on the other side
(Fig. 1). At variance with previous studies, we introduce two distinct
pulling modes: (A) $\mathbf{f}_1=\mathbf{f}$ and
$\mathbf{f}_2=\mathbf{0}$, i.e. one of the two strands is stretched
and the other one is left free, DNA unpeeling; (B)
$\mathbf{f}_1=\mathbf{f}_2=\mathbf{f}/2$, i.e. both strands are
stretched by independent, but equal, forces, stretched DNA
denaturation. Each strand will be considered at some coarse grained
level and modeled as a self-avoiding polymer.
Strand complementarity is taken into account by allowing
a pair of monomers in different strands to be bound with an energy
$-\epsilon<0$ only if their monomer index along the two strands is
equal~\cite{Poland}. Monomer sizes (length of 1 bp) and persistence
lengths are here the only parameters controlling the microscopic
properties of dsDNA and ssDNA. We choose monomer sizes in two
fundamentally distinct ways. If we assume dsDNA exists also in the
stretched S-form, monomer sizes are then equal for both ssDNA and
dsDNA (cases $A_0\ B_0$). If not, dsDNA monomer size, from the B-form,
is smaller than for ssDNA (cases $A_1\ B_1$). The values of monomer
sizes and persistence lengths mostly impact on the large and small
force behaviour of the system respectively.

From now on $T$ will be measured in units of $k_B/\epsilon$ ($k_B$ is
the Boltzmann constant). Unlike all previous work, we perform
numerical simulations finding a phase diagram valid in the whole $T$
range. We can thus discover that models A and B behave amazingly
differently, both qualitatively and quantitatively.

We begin with heuristic arguments to determine the critical force,
$f_c(T)$, separating the zipped phase from the unzipped/melted one.
At $f=0$ we have the standard melting phase transition occurring
\cite{Poland,Carlon} at $T=T_m$, i.e. $f_c(T_m)=0$. For case $A_0$,
unpeeling with same monomer size for ss and dsDNA, at large $f$ the
(almost) completely stretched strand acts as a 1D substrate for the
adsorption of the free (un-stretched) strand. In this case the only
allowed conformation are Y-like configuration with no bubbles. An
energy-entropy argument leads to a vertical asymptote for $f_c(T)$ at
$T_a=\epsilon/s_s$, where $s_s$ is the entropy per monomer of a single
DNA strand~\cite{note_hanke}. This argument is fully confirmed by
numerical results coming from 3D Monte-Carlo simulation of model (A),
i.e. two paired DNA strands, one of which is under a stretching force
of modulus $f$ (see Fig. 2, with typical snapshots from the
simulations shown as well).  Each strand is modelled by a
self-avoiding chain, made up by $N$ beads of size $\sigma$ ($\sim 1$
nm, the size of ssDNA), connected by springs.  To model homologous
base pairing, we chose a truncated Lennard-Jones potential, with
minimum value $-\epsilon$, attained when the corresponding beads are
at a mutual separation $\sigma$. The bonds between successive beads
are harmonic springs with minimal and maximal elongations equal to 0.7
$\sigma$ and $\sigma$ respectively (the maximum elongation is then
$(N-1)\sigma$).  The simulations were performed by proposing local
deformations of the chain. We used multiple Markov chains~\cite{mmc}
to improve sampling efficiency at small $T$ or large $f$, and
reweighted the simulation data prior to analysis according to the
Ferrenberg-Swedsen algorithm~\cite{ferrenberg}.  The critical points
were estimated via the peaks of the specific heat.  Our data suggest a
first order transition at all $f\ne 0$.

\begin{figure}
\centerline {\includegraphics[width=5.cm]{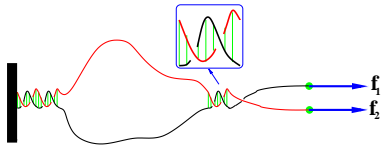}}
\caption{Schematics of our model. A dsDNA molecule
is anchored at one of its ends, whereas two pulling forces stretch each 
strand at the other ends. At a finite $T$ bubbles may appear
in typical DNA configurations as in the figure.}
\end{figure}

The phase diagram found numerically (Fig. 2) shows that there is as
expected a vertical asymptote. At small forces,
the unpeeling pulling mode stabilises the zipped dsDNA, so that the
critical temperature, $T_c(f)$, increases with $f$ (this agrees with a
thermodynamic analysis of DNA overstretching data
\cite{rouzina,rudnick}). 

To study model $B_0$, we have performed our simulations in the case in
which both DNA strands are under the action of the same force
($f_1=f_2=f/2$). As bubbles (see Fig. 1) are stretched out, the two
strands are on average closer to each other with respect to the $f=0$
case and one may argue that the dsDNA should be stabilised by $f$.
Another expectation is that at very large $f$, both strands become
straight in the same direction and the zipped phase is again
stabilised since base pairing does not lead to further entropy
cost. Both these expectations are confirmed by the Monte-Carlo
simulations (see Fig. 2, dotted line): the region of stability of the
zipped phase, in the $(T,f)$ plane, increases, and the denaturation
temperature appears to be a monotonically increasing function of $f$.
The phase diagrams for both models $A_0$ and $B_0$ are somewhat
surprising as at low $T$ (and indeed at any $T$ for model $B_0$) DNA
can stay zipped even at large $f$. This would however be compatible
with the interpretation of overstretched DNA as a ds form different
from B-DNA~\cite{Clausen-Schaumann00}. The key point is that monomer
sizes of S-DNA and ssDNA are roughly the same. The overstretching
transition (B-DNA to S-DNA) takes place in Fig. 2 at $65$pN (the energy
scale $\epsilon$ was set to $2.7$ Kcal/mol, by assuming $T_m=70\
{}^o$C). For unpeeling, overstretching can be followed by further
unzipping to ssDNA, whereas this is not possible when pulling both
strands.  A second elongation transition is indeed observed in some
cases~\cite{Cocco,Geissler}. Our model also suggests that at higher
$T$, and when pulling just one strands, overstretching should change
to a melting transition, even assuming the existence of S-DNA. The
picture emerging from our results might thus reconcile both
interpretations of overstretched DNA.

\begin{figure}
\centerline {\includegraphics[width=5.5cm]{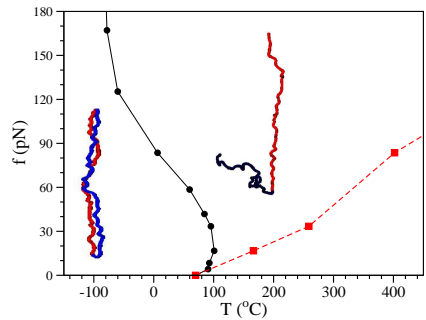}}
\caption{Solid line: $f-T$ phase diagram in the continuum limit for a
model with two self-avoiding chains kept together by a truncated
Lennard-Jones potential (with minimum $-\epsilon$), and a force
stretching one of the strands. There is a first order transition
separating the zipped from the unzipped phases. Dashed line: same for
a force stretching both strands. The temperature range includes
unphysical values to show the whole theoretical diagram.}
\end{figure}

It is instructive to compare the phase diagrams found with the
Monte-Carlo simulations to those of an exactly solvable model for DNA
melting under a stretching force. In these simplified models the
two strands are mutually avoiding directed walks along the [11]
diagonal of a 2D lattice (Fig. 3a). When two beads are in contact they
gain a pairing potential $\epsilon$, and one end of the chain is fixed at
the origin for both strands, while at the other end there are 
stretching forces, $f_1$ and $f_2$ (Fig. 3b). The case with $f_2=0$
corresponds to our model $A_0$. 

One can write down a recursion for the partition function 
of a system with only one degree of freedom, the open end distance $x$. 
If the strands are subjected to two forces $f_1$ and $f_2$ 
(e.g. identified by their projection along the positive [-1 1] vector,
Fig. 2a), then these are:
\begin{eqnarray}\label{recursions1}
Z_{N+1}(x) & = & Z_N(x+1){y_2}/{y_1}+Z_N(x-1){y_1}/{y_2} \\ \nonumber
& + & Z_N(x)\left(y_1y_2+{1}/{(y_1y_2)}\right) \qquad {\rm for} \, x>0
\\ \label{recursions2} Z_{N+1}(0) e^{-1/T}& = & Z_N(1){y_2}/{y_1} \\
\nonumber & + & p Z_N(0)\left(y_1y_2+{1}/{(y_1y_2)}\right) .
\end{eqnarray}
In Eq.s \ref{recursions1},\ref{recursions2} $Z_N(x)$ indicates the
partition function of two walks with open end distance $x$,
$y_1=e^{\beta f_1}$, $y_2=e^{\beta f_2}$, and $p$ is a parameter which
controls the weight of dsDNA segments and which we temporarily set to
1, implying that both monomer sizes and persistence lengths are equal
in both ss and ds phases. The phase diagrams for models $A_0$ and
$B_0$ can be found with e.g. a generalisation of the methods used in
Ref. \cite{dna_unzipping3} and are plotted in Fig. 3b.

Fig. 3b shows that the phase diagram is qualitatively similar to the
one already found in 3D Monte Carlo simulations in the continuum. The
presence of a vertical asymptote for unpeeling, at
$T=T_a=\frac{1}{\log{2}}$, the transition temperature for polymer
adsorption on a wall, is confirmed. Interestingly, while unpeeling, at
$f\ne0$ is first order, stretched denaturation is second order, as can
be seen from the plots of average percentage of zipped bases
($\theta$, Fig. 3c), and of the average open end distance per monomer
($\langle x\rangle$/N, Fig. 3d). (The unpeeling phase diagram has been
found, in another context, in~\cite{somen}.)  Simulation data show
consistently a much smoother transition when pulling both strands. On
the other hand, in the case of DNA unpeeling there are some
differences between Monte Carlo simulations and exact results, most
notably the shape of the phase diagram close to $T_m$, as the critical
temperature {\em decreases with $f$}.

\begin{figure}
\centerline {\includegraphics[width=6.5cm]{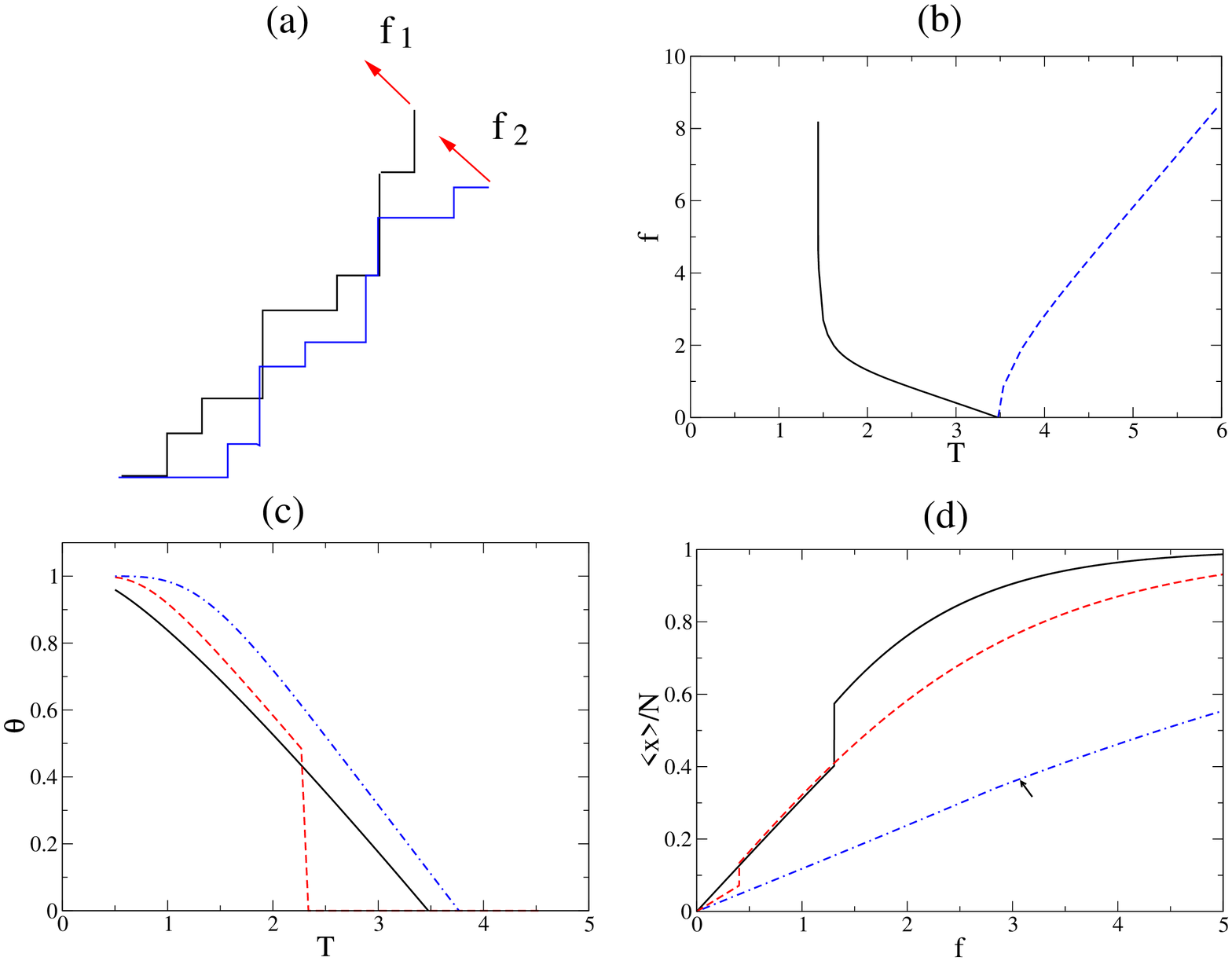}}
\caption{(a) Geometry for the exact calculations.
(b) Phase diagram for unpeeling (solid line) and stretched denaturation
(dashed line). (c) Fraction of zipped bases as a function of
$T$ for $f=0$ (solid line), for DNA unpeeling, and $f=1$ (dashed line), 
for stretched denaturation and $f=2$ (dot-dashed line).
(d) End-to-end distance of the pulled strand per monomer as a function of $f$ 
for DNA unpeeling and $T=2$ (solid line), $T=3$ (dashed line) and 
for stretched denaturation and $T=4$ (dot-dashed line, with the arrow
denoting the transition).}
\label{exact_models}
\end{figure}

Interestingly, an expected difference concerns the order of the
melting transition at zero force, which is thought to be first order in 
3D~\cite{Carlon}, and is second order in the directed model. Another 
realistic feature that the simulated model reproduces
is the larger effective persistence length of the ds chain with
respect to the ss one, at variance with the lattice model.

To gain more insight, we present a Flory-like argument to rationalise
the $f\to 0$ behaviour of the phase diagram for DNA unpeeling. 
We consider both cases $A_0$ and $A_1$, in which
the effective length of dsDNA segments~\cite{noteK} is larger
($\ell_d \sim 300$ bp $\sim 100$ nm for B-DNA)
with respect to that of ssDNA ones ($\ell_s
\sim 8$ bp $\sim 4$ nm). The monomer sizes are $a_d$ for dsDNA and
$a_s\simeq 0.56$ nm for ssDNA, and may differ. Let us
consider $T\simeq T_m$ and a small force $f\ll1$.  If we disregard
denaturation bubbles (considering only Y-shaped configurations),
the energy of two DNA strands with $m=\theta N$ open bp, under the
action of an unpeeling force $\mathbf{f}$, is $E=-\epsilon (N-m) -
\mathbf{f}\cdot\mathbf{r}$, where $\mathbf{r}$ is the position of the
un-anchored end of the stretched strand.  We may expect the
probability density, $P(\mathbf{r}|N,m)$ for the stretched strand to
be described by a Gaussian approximation, in $D$-dimension,
\begin{eqnarray}\label{G}
  P(\mathbf{r}|N,m)&=&\left(2\pi\sigma^2(\theta)N\right)^{-\frac{D}{2}}
    e^{\{-\frac{r^2}{2\sigma^2(\theta)N}\}} \\ \sigma^2(\theta)
    &\propto& (1-\theta) a_d^2\ell_d + \theta a_s^2\ell_s
\end{eqnarray}
The total number of Y-like configurations is given by, modulo power law 
corrections,
$C(N|m)\propto \exp\{\frac{(N-m)}{\ell_d}s+\frac{m}{\ell_s}s\}$
where the entropies per monomer of the double and single stranded DNA have 
been estimated to be $s/\ell_d$ and $s/\ell_s$ respectively.
The total entropy of Y-like configuration with a fixed $\mathbf{r}$ is 
estimated as $S=\log\{C(N|m)P(\mathbf{r}|N,m)\}$.
Minimizing the free energy ${\cal F}=E-TS$ with respect to $\mathbf{r}$ 
and $\theta$, we obtain the following form for the critical force, $f_c(T)$,
\begin{equation}\label{fc}
    f_c^2(T)=\frac{T-T_m}{T_m}\frac{2T}{\ell_da_d^2-\ell_sa_s^2},
\quad T_m=\frac{\epsilon\ell_d\ell_s}{s\left(2\ell_d-\ell_s\right)}.
\end{equation}
Thus if $\ell_da_d^2>\ell_sa_s^2$, as in reality, $T_c$ increases with
$f$. Notice that $f_c(T)\sim (T-T_m)^{1/2}$ at $T\sim T_m^+$
(self-avoidance would change the exponent from $1/2$ to $\nu\simeq
0.588$, although this trend is not discernible in Fig. 2 due,
possibly, to the distance from the thermodynamic limit).  The behavior
of $f_c$ given by eq.(\ref{fc}) is only valid for $T\sim T_m$.  Away
from $T_m$ this transition line has to turn left and e.g. approach a
vertical asymtote for $a_s=a_d$. This is indeed the shape of the phase
diagram obtained with Monte-Carlo simulations for model $A_0$ with
$a_s=a_d$, as $\ell_d>\ell_s$ due to the interaction. Why is the dsDNA
phase instead {\em destabilised} by $f$ in the exact model (fig. 3b)?
A plausible reason is that the argument we have just proposed neglects
denaturation bubbles, and thus predicts $T_m=T_a$ for all forces, when
$\ell_d=\ell_s$, $a_d=a_s$.  Bubbles are unimportant for the continuum
simulations, as the melting is first order for self-avoiding chains,
but are crucial for the exact model, as the transition is second order
and there are $\sim N^{1/2}$ bubbles at the
transition~\cite{dna_unzipping3}. Bubbles increase $T_m$ and the
stability of the zipped phase at $f=0$, and might indeed cause the
small force reentrance to disappear.

We now turn to models $A_1$ and $B_1$, assuming that dsDNA can only be
in the B-form. It is then possible to generalize our exact results by
introducing different values $\xi=\ell_d/\ell_s=25$ and
$a=a_d/a_s=0.6$ for Kuhn lengths and monomer sizes of B-DNA and
ssDNA. To incorporate these elements into our models, we may weigh
dsDNA segments in Eq. \ref{recursions2} as follows:
\begin{equation}\label{overstretching}
p(\beta,f,\xi,a)=
{\left[2{{\rm cosh}\left(\beta f a\xi\right)}\right]^{1/\xi}}/
{2{{\rm cosh}\left(\beta f \right)}}
\end{equation}
The resulting phase diagrams are shown in Fig. 4. The most striking
difference with respect to the phase diagrams in Figs. 2 and 3 is that
there is now a bound region for the zipped phase, and at large force
the DNA is always ss and denatured, in agreement with the
interpretation of overstretched DNA as melted DNA. A second difference
is that close to the denaturation transition the shape of the critical
line is affected by denaturation bubbles as well. As a result for
directed walks dsDNA is stabilised by a force only when both strands
are stretched. When the transition is first order (DNA unpeeling),
then we can derive $df_c(T)/dT=-\delta S_{u,z}/ \delta x_{u,z}$, where
$\delta S_{u,z}$ and $\delta x_{u,z}$ are the changes in entropy and
elongation (along the direction of $\mathbf{f}$) between the zipped
and the unzipped phases~\cite{rouzina}. Close to $T=0$ for DNA
unpeeling both these quantities are positive, hence there is no
reentrance, whereas close to the melting temperature $\delta
S_{u,z}>0$ but the sign of $\delta x_{u,z}$ may vary.

To conclude, we presented Monte-Carlo simulations and exact results to
determine the phase diagram of dsDNA melting in the presence of an
external force, stretching one or both strands -- DNA unpeeling or
stretched DNA denaturation.  Our results show that these two cases,
which may often be difficult to distinguish in single molecule
experiments, lead to strikingly different results, both qualitatively
and quantitatively. Contrarily to previous claims, our results suggest
that DNA unpeeling is a first order phase transition.  We have
discussed how to introduce in our models the realistic persistence
lengths and monomer sizes of ds and ssDNA. This can be done in
different ways according to what is assumed about the nature of the
overstretched DNA state.  Even if we assume dsDNA can exist in the
S-DNA form we have shown that overstretching transition to a melted
ssDNA is possible {\it when one strand is being pulled}.  According to
our calculations, we may infer the phase diagram of a realistic model,
taking these effects as well as self-avoidance, fully into account. We
predict that it should (i) display no ``reentrance'' for low $T$, and
(ii) show dsDNA stabilisation by force, close to
$T_m$~\cite{realistic_argument}.\\

\begin{figure}
\centerline {\includegraphics[width=6.cm]
{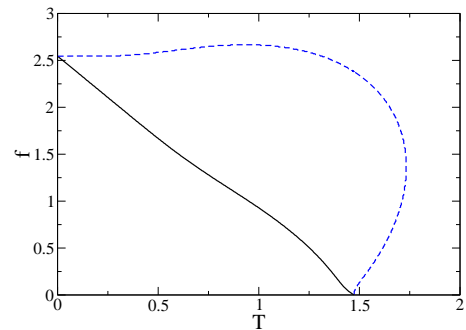}}
\caption{$(T,f)$ phase diagrams for model $A_1$ (solid line) and 
$B_1$ (dashed line), with $a_d=0.34$, $l_d=50$ nm, and $a_s=0.56$, 
$\ell_s=2$ nm for ds and ss DNA respectively. }
\end{figure}


\begin{thebibliography}{99}
\bibitem{Marko} J. F. Marko, E. D. Siggia, {\it Macromolecules}
{\bf 28}, 8759 (1995); M. Rief, H. Clausen-Schaumann, H. E. Gaub,
{\it Nat. Struct. Biol.} {\bf 6}, 346 (1999).
\bibitem{Essevaz} B. Essevaz-Roulet, U. Boeckelmann, F. Heslot,
Proc. Natl. Acad. Sci. USA {\bf 94} 11935 (1997).
\bibitem{rouzina} M. C. Williams {\em et al.}, 
{\it Biophys. J.} {\bf 80}, 1932 (2001);
I. Rouzina, V. A. Bloomfield, {\it Biophys. J.}
{\bf 80}, 882 (2001).
\bibitem{Clausen-Schaumann00} H. Clausen-Schaumann {\em et al.},
{\it Biophys. J.} {\bf 78}, 1997 (2000). 
\bibitem{Cocco} S. Cocco {\em et al.}, {\it Phys. Rev. E} {\bf 70}, 
011910 (2004).
\bibitem{Nelson} C. Storm, P. C. Nelson {\it Phys. Rev. E} {\bf 67}
051906 (2003).
\bibitem{Geissler} S. Whitelam, S. Pronk, P.~L. Geissler, {\it
Biophys. J.} {\bf 94}, 2452 (2008).
\bibitem{dna_unzipping1} S. M. Bhattacharjee, {\it J. Phys. A}
{\bf 33}, L423 (2000).
\bibitem{dna_unzipping3}
D. Marenduzzo, A. Trovato, A. Maritan, {\it Phys. Rev. E} {\bf 64}, 031901
(2001); 
D. Marenduzzo {\em et al.}, {\it Phys. Rev. Lett.} {\bf 88}, 028102 (2002).
\bibitem{stretching} D. Marenduzzo {\em et al.}, {\it Phys. Rev. Lett.} 
{\bf 90}, 088301 (2003).
\bibitem{kumar} S. Kumar {\em et al.}, 
{\it Phys. Rev. Lett.} {\bf 98}, 128101 (2007).
\bibitem{somen} R. Kapri, S. M. Bhattacharjee, e-print arXiv:0803.3440;
R. Kapri, e-print arXiv:0803.3443.
\bibitem{Hanke} A. Hanke, M. G. Ochoa, R. Metzler, {\it Phys. Rev. Lett.}
{\bf 100}, 018106 (2008).
\bibitem{rudnick} J. Rudnick, T. Kuriabova, e-print arXiv:0709.3846.
\bibitem{Poland} D. Poland, H. A. Scheraga, {\it J. Chem. Phys.}
{\bf 45}, 1456 (1966).
\bibitem{Carlon} Y. Kafri, D. Mukamel, L. Peliti, {\it
Phys. Rev. Lett.}  {\bf 85}, 4988 (2000); E. Carlon, E. Orlandini,
A. L. Stella, {\it Phys. Rev. Lett.} {\bf 88}, 198101 (2002).
\bibitem{note_hanke} This is in contrast with the phase diagram in\cite{Hanke},
which at low $T$, shows a reentrant transition line. This is due to the fact 
that, as noted by the authors, 
assuming a Gaussian probability distribution for the ends of the chains 
is not valid at low $T$.
\bibitem{mmc} M. C. Tesi {\em et al.}, {\it J. Stat. Phys.} {\bf 82}, 
155 (1996).
\bibitem{ferrenberg} A. M. Ferrenberg, R. H. Swendsen,
{\it Phys. Rev. Lett.} {\bf 63}, 1195 (1989).
\bibitem{noteK} The effective segment length (or Kuhn length) is twice the
persistence length for a freely jointed chain.
\bibitem{realistic_argument} No ``reentrance'' means here that
$df_c(T)/dT$ at $T=0$ is $\le 0$. To see why there is dsDNA
stabilisation, consider unpeeling: it is first order, thus the
Flory-like argument leading to Eq. \ref{fc}, and predicting
stabilisation for realistic values of $\ell_s$, $a_s$, $\ell_d$ and
$a_d$, holds. Stretched denaturation enhances stabilisation
(Figs. 2-4).


\end{thebibliography}
 \end{document}